\documentclass[fleqn,10pt]{wlscirep}
\usepackage[utf8]{inputenc}
\usepackage[T1]{fontenc}
\usepackage{comment}
\usepackage{float}

\usepackage{graphicx}
\usepackage{makecell}
\usepackage{xcolor}
\usepackage{multirow}
\usepackage{threeparttable}
\usepackage{braket}
\usepackage{mathtools}
\usepackage{amsmath}
\usepackage{caption}
\usepackage{chemfig}
\usepackage{chemformula}
\usepackage{subcaption}
\usepackage[section]{placeins}

\title{Single-Photon Double Ionization of Ozone}

\author[1]{Veronica Daver Ideböhn}
\author[2]{Antoine Gloriod}
\author[1]{Richard J. Squibb}
\author[1]{Andreas Hult Roos}
\author[1]{Nihar Ranjan Behera}
\author[1]{Ishita Kanungo}
\author[1]{Elias Gustafsson}
\author[1]{Simon Gällblad}
\author[1]{Saga Berglund}
\author[1]{Emelie Olsson}
\author[3]{Gunnar Öhrwall}
\author[4]{Gunnar Nyman}
\author[5]{John M. Dyke}
\author[6]{John H.D. Eland}
\author[2,+]{Majdi Hochlaf}
\author[1,++]{Raimund Feifel}

\affil[1]{University of Gothenburg, Department of Physics, Origovägen 6B, 412 58 Gothenburg, Sweden}
\affil[2]{Université Gustave Eiffel,  COSYS/IMSE, 5 Bd Descartes, 77454 Champs sur Marne, France}
\affil[3]{MAX IV Laboratory, Lund University, Box 118, SE-22100 Lund, Sweden}
\affil[4]{University of Gothenburg,  Department of Chemistry and Molecular Biology, Box 462, 405 30 Gothenburg, Sweden}
\affil[5]{School of Chemistry, University of Southampton, Highfield, Southampton SO17 1BJ, United Kingdom}
\affil[6]{Oxford University, Department of Chemistry, Physical and Theoretical Chemistry Laboratory, South Parks Road, Oxford OX1 3QZ, United Kingdom}
\affil[+]{majdi.hochlaf@univ-eiffel.fr}
\affil[++]{raimund.feifel@physic.gu.se}

\begin{abstract}
Ozone (O$_3$) is a triatomic molecule of central importance in the chemistry and physics of the Earth’s and other planetary atmospheres. Beyond its environmental significance, a detailed understanding of the electronic structure and ionization dynamics of ozone is essential for modeling atmospheric, ionospheric, and astrochemical processes. 
In the present work, we substantially extend the experimental and theoretical characterization of ozone into the regime of valence double photoionization. Using HeII-$\alpha$, HeII-$\beta$, and higher-energy vacuum ultraviolet radiation in combination with a versatile multiple charged-particle correlation detection technique, we report the first single-photon valence double ionization electron spectrum of O$_3$.
To interpret the experimental observations, we mapped the lowest potential energy surfaces of O$_3^{2+}$ employing post-Hartree--Fock multi-configurational-interaction methods, and computed with high accuracy the energetics of the relevant dissociation channels. Our results demonstrate that dissociative double ionization of ozone produces electronically excited O$^+$ fragments in addition to the ground-state O$_2^+$ + O$^+$ dissociation pathway, revealing a richer fragmentation dynamics than hitherto recognized.

\end{abstract}

\begin{document}
\newcommand{\ozone}{$\mathrm{O_3}$}
\newcommand{\ozonep}{$\mathrm{O_3^+}$}
\newcommand{\ozonepp}{$\mathrm{O_3^{2+}}$}

\flushbottom
\maketitle

\thispagestyle{empty}

\section*{Introduction}

Ozone (O$_3$) is a triatomic molecule of central importance in the chemistry and physics of the Earth’s and other planetary atmospheres.\cite{calder2025abiotic} In the stratosphere, ozone forms a protective layer that absorbs much of the energetic ultraviolet (UV) radiation of the sun, preventing biologically harmful wavelengths from reaching the Earth’s surface and thus protecting living organisms including mankind from DNA and tissue damage. In contrast, high concentrations of ozone in the lower atmosphere constitute a major air pollutant, causing oxidative stress in plants and animals and contributing to respiratory and cardiovascular disease in humans. The dual environmental role of ozone has rendered it a key topic in atmospheric science for many decades. Following the discovery of the Antarctic ozone hole \cite{farman1985ozonehole}, intense research efforts have been devoted to understanding both the depletion mechanisms of stratospheric ozone and the formation of tropospheric ozone. These concerns are reflected in international agreements such as the Kyoto\cite{kyoto1997} (1997) and Gothenburg protocols\cite{gbg1999} (1999, revised in 2012), which aim to mitigate the adverse effects of ozone imbalance on climate and health.

From a fundamental scientific perspective, detailed knowledge of ozone’s electronic structure, energetics, and reactivity across its various states of charge and excitation is essential for modeling atmospheric and astrochemical processes. The properties of neutral ozone have been extensively investigated using a range of spectroscopic techniques, providing accurate information on its vibrational structure, potential energy surfaces, and symmetry properties. As can be found in chemistry textbooks, the neutral ozone molecule adopts a bent geometry of $\mathrm{C}_{2v}$ symmetry, and its ground-state electronic configuration is given by (see e.g. Ref.\cite{mahla2025photoionization}): 

\((1a_1)^2(1b_2)^2(2a_1)^2(3a_1)^2(2b_2)^2(4a_1)^2(5a_1)^2(3b_2)^2(1b_1)^2(4b_2)^2(6a_1)^2(1a_2)^2\)


Singly positively charged ozone (O$_3^+$) has also been the subject of considerable attention. Its energy levels and vibrational structure have been characterized by vacuum ultraviolet (VUV) photoelectron spectroscopy\cite{dyke1974vacuum,brundle1974he,frost1974high,katsumata1984photoelectron,mocellin2001experimental}, threshold photoelectron spectroscopy\cite{couto2006threshold}, and high-resolution pulsed-field-ionization zero-kinetic-energy (PFI-ZEKE)\cite{willitsch2005high} techniques. These studies have elucidated the fine structure of the lowest cationic states and revealed strong vibronic coupling and state mixing, reflecting the rich multi-configurational nature of ozone. Core-level photoelectron spectroscopy has further provided site-specific binding energies for the terminal and central oxygen atoms\cite{wiesner2003electronic}, offering insight into the localization of charge and the dissociation dynamics following core excitation.

By contrast, doubly ionized ozone (O$_3^{2+}$) remains much less well understood. From a theoretical standpoint, the removal of two electrons from this highly correlated triatomic system is expected to result in multiple electronic states with complex potential energy surfaces and strong coupling between electronic and nuclear degrees of freedom. Experimentally, only a few studies have addressed so far the double ionization of ozone, most notably those based on electron-impact ionization combined with ion–ion coincidence detection\cite{newson1995electron}. These experiments identified a threshold of 34.4 eV for forming the lowest-lying dicationic state, which was observed to dissociate into O$^+$ and O$_2^+$ fragments. Computations suggest that this threshold corresponds to the adiabatic ionisation energy of $\mathrm{O_3}$ (of 34.7 – 35.0 eV), whereas the vertical ionization energy is found higher (at $\sim$ 35 – 36 eV) because of unfavourable Franck-Condon factors upon doubly ionising ozone (from bent to linear transition). \cite{champkin1999dissociation,pyykko1989ab,pyykko1990ab} Besides, the electron-impact dissociation of $\mathrm{O_3^+}$ is shown to produce $\mathrm{O^+}$ + $\mathrm{O_2^+}$ and $\mathrm{O^+}$ + $\mathrm{O^+}$ + O fragments in addition to $\mathrm{O^+}$ + $\mathrm{O_2}$ and $\mathrm{O^+}$ + O + O channels. The formers are most likely due to the unimolecular decomposition of the intermediate $\mathrm{O_3^{2+}}$ dication formed upon ionising $\mathrm{O_3^{+}}$ cation. \cite{belic2015electron,PhysRevA.82.062715} So far, the $\mathrm{O_3^{2+}}$ parent ion was not observed, being too short lived to reach the detector. 
For explanation, Price and co-workers showed that the dicationic potential energy surfaces of singlet – triplet spin cross in the Franck – Condon region upon doubly ionising \ozone. The triplet is, however, dissociative and leads to $\mathrm{O_2^{+}}$  + $\mathrm{O^{+}}$  products and thus prevents on measuring long-lived $\mathrm{O_3^{2+}}$ ions by impact ionisation of $\mathrm{O_3}$. \cite{champkin1999dissociation} While such measurements provided already some valuable energetic benchmarks, the valence double ionization electron spectrum of ozone has not previously been measured, and little is known about its photon-induced fragmentation dynamics or the stability of its excited dicationic electronic states.

In the present work, we extend substantially the experimental and theoretical characterization of ozone further into the realm of double ionization. Using HeII-$\alpha$, HeII-$\beta$ and higher energy vacuum ultraviolet radiation and a versatile multiple charged-particle correlation detection method, we have recorded the first single-photon valence double ionization electron spectrum of O$_3$. The choice of HeII-$\beta$ radiation is particularly relevant, as helium emission is a dominant component of the solar spectrum, rendering these results directly applicable to photoionization and ion–molecule processes in the Earth’s ionosphere and in other planetary atmospheres. The measured double ionization energies of O$_3$ provide new benchmarks for theory and enable an improved understanding of the energy available for the dissociation and chemical reactivity of ionic O$_3$. For interpretation of the experimental observations, we mapped the lowest potential energy surfaces of O$_3$ dications using post-Hartree-Fock multiconfiguration interaction approaches. We also computed with high accuracy the energetics of $\mathrm{O_3^{2+}}$ and its dissociation channels. With these results, we aim to elevate our fundamental understanding of ozone’s electronic structure well beyond single-ionization and to contribute to a more comprehensive picture of the energy available for ion-molecule reactions of dicationic O$_3$ in the Earth’s ionosphere and extraterrestrial planetary atmospheres.

\section*{Experiments}

The experiments were carried out in our laboratory at the University of Gothenburg and at the synchrotron radiaton facility MAX IV in Lund using time-of-flight photoelectron–photoelectron (TOF-PEPECO) and photoelectron–photoelectron–photoion–photo
ion–coincidence (TOF-PEPEPIPICO) set-ups based on magnetic bottles, as described previously \cite{eland2003complete,PEPItechnique}. In Gothenburg, the TOF-PEPECO measurements employed a unique 5.6\,m instrument consisting of a conical $\sim$1\,T magnet in the light–matter interaction region coupled to a solenoid running along the entire flight tube and producing a magnetic field of a few mT. The sample was introduced into the vacuum chamber through a stainless-steel needle, forming an effusive jet in the interaction region. A pulsed helium discharge lamp provided radiation at a repetition rate of about 4.4\,kHz, with the photon energy selected using an ellipsoidal grating with a groove density of 1100\,lines/mm. This enabled measurements at the discrete photon energies 21.22, 40.81, and 48.37\,eV (He\,I-$\alpha$, He\,II-$\alpha$, and He-\,II$\beta$, respectively) in a small interaction volume. The 5.6\,m instrument has a collection–detection efficiency of $\sim$ 40 \% and a nominal resolving power $E/\Delta E$ of $\sim$ 120. At the synchrotron radiation facility MAX IV, operated in single bunch mode, the beamline FlexPES was utilized which is equipped with a mechanical chopper system to reduce the light frequency further from 3.125 MHz to about 94 kHz to allow for sufficient time for all the electrons to reach the detector before the next radiation pulse acts. The photon energy used in the measurements was 56 eV and the data were collected with a 2.2 m magnetic bottle with a resolving power of $E/\Delta E$ of $\sim$ 50, while otherwise  being very similar to the 5.6 m instrument. 

The TOF-PEPEPIPICO measurements were performed in Gothenburg using the 2.2\,m instrument in which the conical magnet was replaced by a hollow ring magnet of reduced field strength. A repeller and pulser plate were installed such that suitable voltages were applied $\sim$50\,ns after ionization to guide the nascent ions into the 0.12\,m ion flight tube in the direction opposite to the electron flight tube, triggered once the electrons had left the interaction region as described in Ref. \cite{PEPItechnique}. The potentials guiding the ions were adjusted according to Wiley–McLaren conditions to achieve both energy and time focusing.\cite{WileyMcLaren} Replacing the conical magnet with a hollow ring magnet and using a shorter flight tube reduces the nominal resolving power of the electrons to $E/\Delta E = 20$ and yields a collection–detection efficiency of approximately 20 \% for ions. The photon energy was selected using an ellipsoidal grating with a groove density of 550\,lines/mm and the repetition rate was about 4.4 kHz. To ensure that electrons and ions recorded within the same time window originate from the same ionization event, the ionization rate is ideally kept below 2 \%. However, because of comparatively low vapor pressure in ozone and thus relatively high background gas contribution in the apparatus, the count rates were chosen to be slightly higher than ideal, leading to some unwanted coincidences. In both the electron-only and electron–ion configurations, a small voltage ($<1$\,V) was applied across the interaction region to enable collection of low–kinetic-energy electrons. 

Ozone was generated using a commercially available ozoniser from C-Lasky and adsorbed onto silica beads in a U-tube maintained at 195\,K using a dry-ice–isopropanol slush bath. The ozone-loaded U-tube was transferred at 195\,K to the spectrometer and kept at 195 K whilst allowed to desorb from the beads while being pumped through the gas-handling system and simultaneously introduced into the spectrometer, providing an ozone spectrum with small amounts of O$_2$ \cite{dyke1974vacuum}, and some air contamination. To monitor sample purity throughout the measurements, valence single-ionization spectra were taken periodically in all the acquisitions, and additional ion spectra were collected in the case of TOF-PEPEPIPICO measurements. The main contaminants identified were helium, water, molecular oxygen, and molecular nitrogen, where the ionized oxygen molecule can be destinguished from dissociated ozone by peak width from kinetic energy release. 

\section*{Theory}

Two types of ab initio computations were carried out in order to provide an interpretation of the experimental features observed upon doubly photoionizing \ozone. We indeed computed the adiabatic (ADIE) and the vertical (VDIE) double ionization energies of O$_3$ and mapped its electronic states lying in the 0 – 8 eV energy range with respect to the electronic ground dicationic state. We used the approaches described in Ref. \cite{hochlaf2026positively} as implemented in the MOLPRO package (version 2024) [2] where the atoms were described using the basis sets \cite{dunning1989gaussian,kendall1992electron,woon1993gaussian} of Dunning and co-workers together with their associated density fitting sets\cite{yousaf2008optimized} while using the F12 approaches. In the following we will briefly describe these computations.

The potentials of O$_3^{2+}$ were mapped using the standard and explicitly correlated versions of the internally contracted multi-reference configuration interaction (MRCI  \cite{werner1988efficient,knowles1985efficient,shamasundar2011new} and MRCI-F12  \cite{shiozaki2011explicitly,shiozaki2011explicitly2,shiozaki2013multireference}) methods on top of state-averaging complete active-space self-consistent field (SA-CASSCF) computations\cite{knowles1985efficient,werner1985second}. In CASSCF, the complete active space is composed by considering the orbitals from HOMO-9 up to HOMO+4 (i.e. all the 9 valence orbitals and the 4 next virtual orbitals) as active. All valence electrons were correlated. All electronic states having the same spin-multiplicity were averaged together using the state averaging procedure of MOLPRO. At MRCI or MRCI-F12, the active space was constructed by considering all single and double excitations from the CASSCF wavefunctions, retaining only those with CI coefficients $\geq$ 0.05. Per spin multiplicity, we asked for 4 states per C$_s$ symmetry component. 

To determine ADIEs and VDIEs of the lowest singlet and triplet states of O$_3^{2+}$, we used the (R)CCSD(T)-F12b/aug-cc-pVTZ (opt) // RCCSD(T)/aug-cc-pVTZ + $\Delta$CV + $\Delta$SR + $\Delta$ZVPE (SP) composite scheme, where “opt” and “SP” denote full geometrical optimisations without constraints (in the Cs point group) and single point computations, respectively. (R)CCSD(T) \cite{hampel1992comparison, knowles1993coupled, knowles2000erratum,deegan1994perturbative} and (R)CCSD(T)-F12  \cite{adler2009local,adler2009local2,knizia2009simplified,adler2007simple} are the standard and the explicitly correlated versions of the coupled clusters with single, double, and perturbative triple excitations. $\Delta$CV, $\Delta$SR, $\Delta$ZVPE correspond to the core-valence, scalar relativistic and zero-point vibrational energy corrections. The later was evaluated after a PBE0/aug-cc-pVTZ geometry optimisation followed by anharmonic frequency calculations. ADIEs of the lowest singlet and triplet states of O$_3^{2+}$ were evaluated as the energy difference between that of the dication and the neutral taken at their respective equilibrium geometries, whereas their VDIEs were determined as the energy difference between that of this dication electronic and the neutral at the equilibrium geometry of the neutral. Then, we located the upper dication electronic states using their relative energies with respect the lowest O$_3^{2+}$ singlet as computed at the MRCI-F12/CASSCF/aug-cc-pVTZ level. These computations were done in C$_{2v}$ point group where we requested 4 states per C$_{2v}$ symmetry for a given spin multiplicity. The expected accuracy is 0.02 eV or better.

\section*{Results and Discussion}

\section{Electron-electron coincidences}
\begin{figure}[H]
    \centering
    \includegraphics[width=0.4\linewidth]{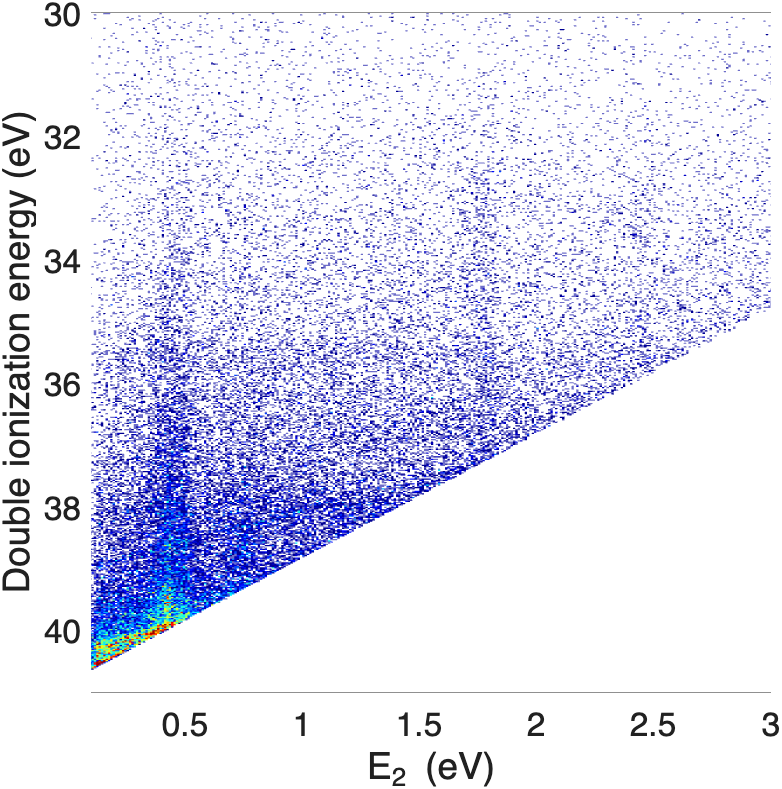}\includegraphics[width=0.4\linewidth]{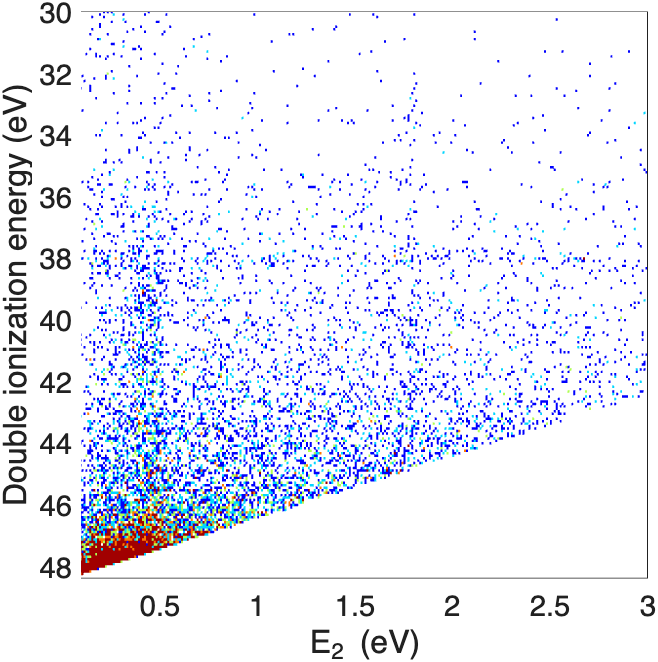}
    \caption{Electron–electron coincidence maps of ozone obtained at 40.81 eV (left) and 48.37 eV (right). The most prominent straight vertical lines just below 0.5 eV, $\sim$ 0.8 eV and $\sim$1.8 eV agree with energies as previously found in related measurements on molecular oxygen, where they were interpreted as arising from autoionization of atomic oxygen. The horizontal lines are from energy sharing double ionization. In the 40.81 eV data, both the broader peak at 35.45 eV and the sharper feature at 38 eV are visible, whereas mainly the 38 eV peak is visible in the 48.37 eV data which comprise fewer total counts.}
    \label{fig:autoionization}
\end{figure}

Figure \ref{fig:autoionization} shows electron–electron coincidence maps of ozone obtained at 40.81 and 48.37 eV, with the electron-pair energy relative to the photon energy - i.e., the double-ionization energy - along the vertical axis, and the kinetic energy $E_2$ of the second arrival electron (out of two) along the horizontal axis. A horizontal line at the double-ionization energy of $\sim$ 38 eV appears as a pronounced feature in these data and reflects a final dicationic electronic state, and in the 41 eV data set, a line at 35.5 eV is also visble. In addition, several straight vertical lines are discernible at both photon energies, with the most prominent one just below 0.5 eV and two additonal ones at $\sim$ 0.8 eV and $\sim$1.8 eV. These features match the autoionization features of atomic oxygen, well known from related experiments on molecular oxygen \cite{feifel2005valence}, where their energies were determined as 0.425 eV, 0.495 eV, 0.754 eV, 0.840 eV, 1.685 eV, and 1.785 eV, respectively.

\begin{figure}[H]
    \centering
    \includegraphics[width=0.85\linewidth]{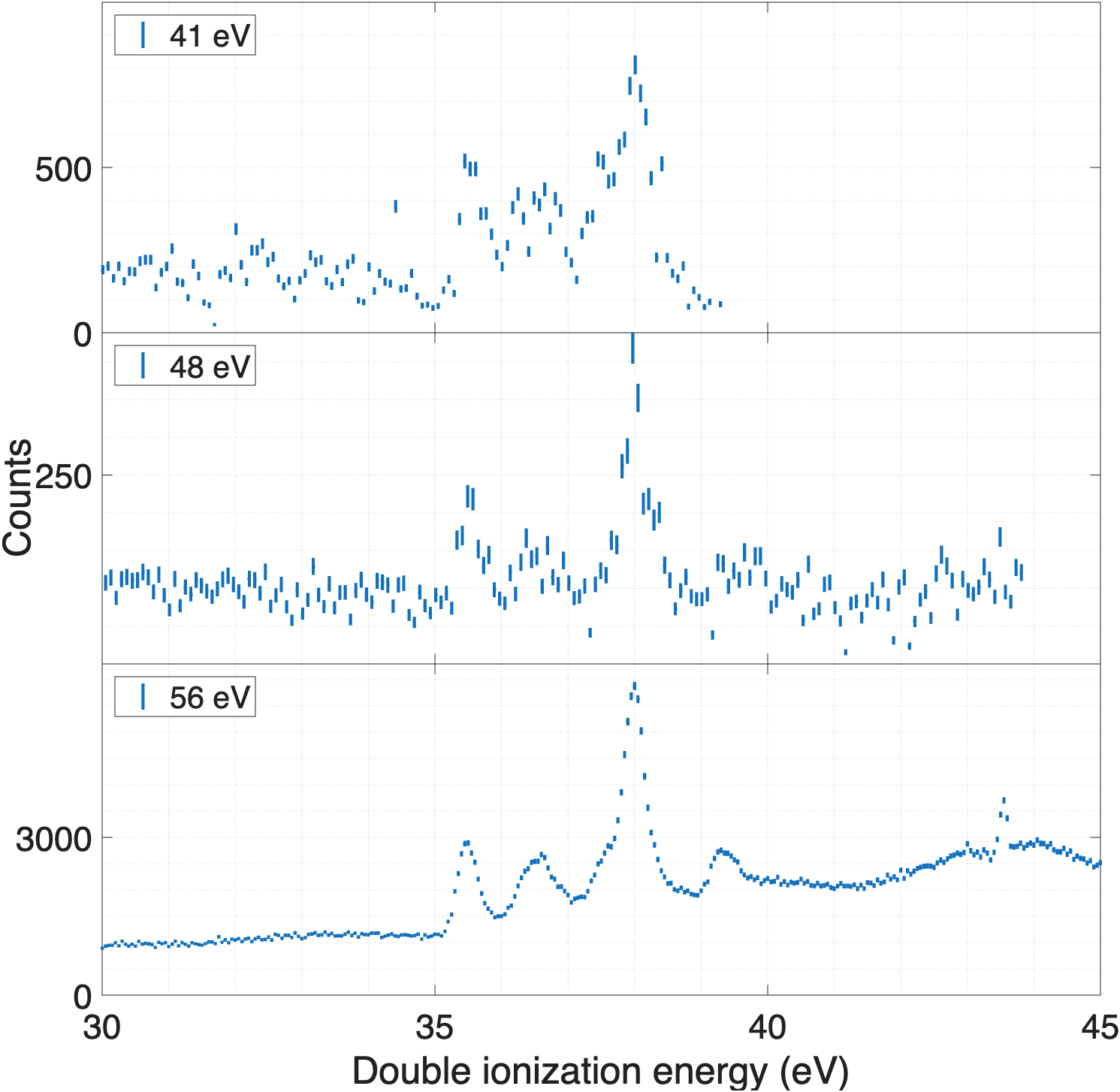}
    \caption{Double ionization electron spectra of ozone taken at 40.81, 48.37 and 56 eV (upper, middle, lower panel respectively). In all spectra, the first, statistically relevant peak is located at the vertical ionization energy of 35.5 eV, followed by a broad feature just above 36 eV. The strongest peak has a vertical ionization energy of 38 eV. Background has been removed from all spectra to emphasize the features. In the upper two spectra this was done thorough a polynomial subtraction and in the 56 eV, electrons with kinetic energy lower than 3.5 and higher than 56-3.5 eV have been excluded. Both methods reduce the effect of autoionization electrons and the natural background increase as double ionization energy approaches the photon energy.}
    \label{fig:ozoneDIE48eV}
\end{figure}

Figure~\ref{fig:ozoneDIE48eV} shows projections of the coincidence maps from Fig.~\ref{fig:autoionization} onto the double-ionization-energy axis, with the addition of a spectrum taken at 56 eV, representing the dicationic electron spectrum of ozone.  The 41 and 48 eV spectra presented have had polynomial background removed, and in the 56 eV spectrum, all electrons with kinetic energy lower than 3.5 and higher than 52.5 eV have been removed, in both instances to enhance the spectral features. For the identified peaks, the energy resolution is approximately 0.1\,eV in the spectrum taken at 48.37\,eV and 0.05\,eV in the 40.81\,eV spectrum. In the 56 eV data, the overall resolving power is just below 0.6 eV for the peaks of interest. In these projections, the previously mentioned atomic autoionization lines have been removed from the data. In the 56 eV data, we see traces of $\mathrm{O_2^{2+}}$ as a shoulder around 37 eV and as a low intensity peak at 43.55 eV.

As seen in Figure \ref{fig:ozoneDIE48eV}, the first peak appears at a vertical ionization energy of 35.5\,eV, followed by a broader feature around 36\,eV, and a sharp peak at 38\,eV. The first peak is consistent with the basic ``rule of thumb'' for double ionization, $\mathrm{DIE} = 2.8 \times \mathrm{IE}$ \cite{tsai1980mass}. According to the present calculations, the lowest predicted doubly ionized electronic state is $\mathrm{X^1\Sigma_g^+}$ with an adiabatic energy of 34.445\,eV and a vertical ionization energy of 35.893\,eV, both in Table \ref{tab:majdistable}. However, calculations show that this state is linear, as seen in Figure \ref{fig:O3_mol}, and lies well outside the Franck–Condon region, and is therefore expected to appear as a broad, weak feature far from the adiabatic double ionization energy in the double ionization electron spectrum. Consequently, the first experimental peak is most likely associated with the lowest triplet state (A$\mathrm{^3B_2}$), for which the adiabatic double ionization energy is computed at 35.513 eV and the vertical double ionization energy is located at 35.603\,eV, both listed in Table \ref{tab:majdistable}. Worth noting that we have similar shape for the valence double ionisation of SO$_2$ populating the $\mathrm{SO_2^{2+}}$ electronic states. \cite{hochlaf2004theoretical} 

\begin{figure}[H]
    \centering
    \includegraphics[width=\linewidth]{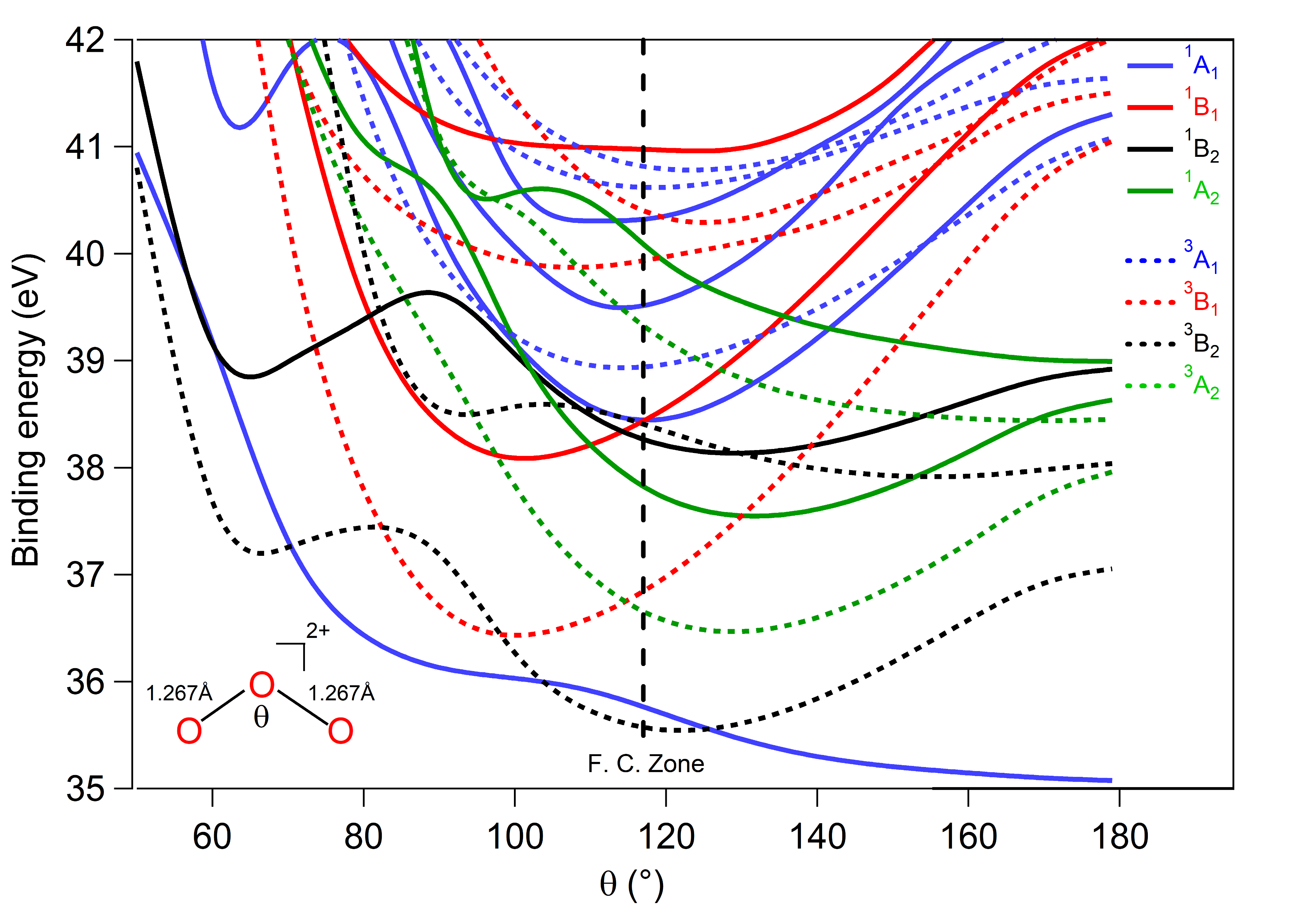}
    \caption{CASSCF/aug-cc-pVTZ one dimensional cuts of the potential energy surfaces of \ozonepp excited electronic states along the in-plane O-O-O angle. The OO distances were set to their CCSD(T)-F12/aug-cc-pVTZ equilibrium values in \ozone ($\mathrm{X^1A_1}$). F.C. denotes the middle of the Franck-Condon zone accessed from \ozone ($\mathrm{^1A_1}$). These potentials are given in energy with respect to that of \ozone ($\mathrm{X^1A_1}$) at equilibrium.}
    \label{fig:angle_theory}
\end{figure}

\begin{figure}[H]
    \centering
    \includegraphics[width=0.5\linewidth]{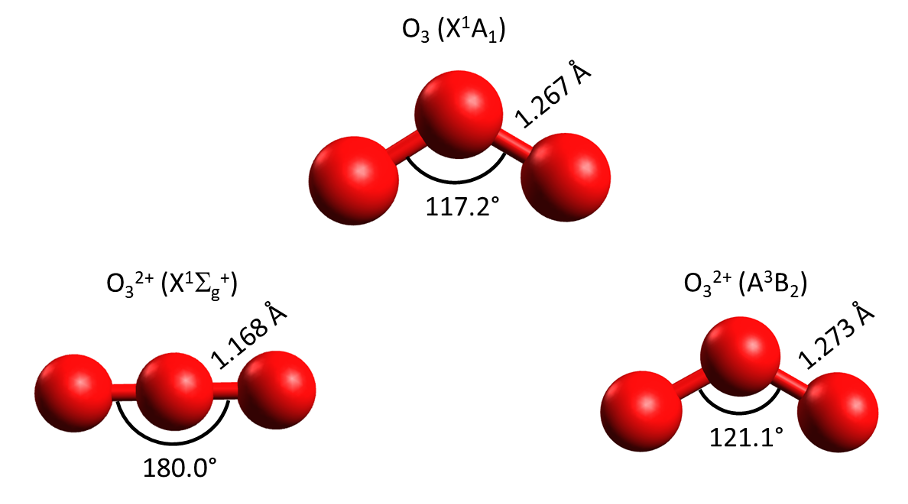}
    \caption{(R)CCSD(T)-F12/aug-cc-pVT(5?)Z optimised structure of the ground state of neutral ozone (top) and the two first electronic state of ozone dication (bottom). The geometrical parameters and spectroscopic term are specified.}
    \label{fig:O3_mol}
\end{figure}

The double ionization electron spectra of O$_3$ recorded at 41, 48 and 56 eV photon energy in Figure \ref{fig:ozoneDIE48eV} display three main features at around 35.5, 36 and 38 eV. However, attributing each of these features to a single electronic state would be an oversimplification. As            clearly evidenced by the potential energy surface cuts presented in Figures \ref{fig:angle_theory} and \ref{fig:rOO_theory}, the electronic structure of $\mathrm{O_3^{2+}}$ is characterized by an exceptionally high density of states, with numerous singlet and triplet states lying within a narrow energy window (36 to 42 eV). These states interact strongly through a variety of non-adiabatic and relativistic mechanisms such as avoided crossings between same symmetry states, possible conical intersections, spin-orbit coupling between different spin states and also Renner-Teller effects along the bending coordinate (cf. Figure \ref{fig:rOO_theory}). With this on our mind, the feature at 35.5 eV can tentatively be assigned as the lowest triplet state, which appears relatively isolated from the other states in the Franck-Condon region at this binding energy. This assignment is supported by two main arguments: the vertical double ionization energy of this state (35.546 eV at MRCI-F12/aVTZ level of theory) is in excellent agreement with the observed peak position, and the close geometry between the one of the dication and the one of the ground state of the neutral ozone. This geometric similarity also implies a favourable Franck-Condon overlap, consistent with the relatively sharp appearance of this feature. However, the overlap of its potential energy curve with the one of the first singlet state, around the Franck-Condon area, may indicate contribution of spin-orbit coupling for this structure. By contrast, the structures between 36 and 42 eV arise from a dense manifold of strongly interacting states whose individual contributions cannot be disentangled from the present spectra alone.


In Figure \ref{fig:angle_theory}, we give an overview of the potentials of \ozonepp  along the in-plane OOO angle, where both OO distances were kept fixed at their values in \ozone($\mathrm{X^1A_1}$). This figure reveals that electronic structure of \ozonepp is characterized by a high density of states, with numerous singlet and triplet states lying within a narrow energy window (35 to 42 eV). These states interact strongly through a variety of non-adiabatic and spin-orbit interactions. Indeed, Figure \ref{fig:angle_theory} shows that several avoided crossings occur between the electronic states having the same space and spin terms even at low energies. Also, spin-orbit couplings between singlets and triplets at their crossings are possible. In addition, all doubly degenerate electronic states at linearity split into two components upon bending the \ozonepp molecule. Both components are coupled by Renner-Teller effect. Besides, Figure \ref{fig:angle_theory} shows that ground state of \ozonepp is located for linear configuration i.e. outside the Franck-Condon (F.C.) zone, whereas the lowest $\mathrm{^3B_2}$ state is located in this zone. Consequently, unfavourable (favourable) Franck-Condon factors (FCFs) are expected for the \ozone ($\mathrm{X^1A_1}$) $\rightarrow$ \ozonepp($\mathrm{X^1\Sigma_g^+}$) (\ozone ($\mathrm{X^1A_1}$) $\rightarrow$ \ozonepp($\mathrm{A^3B_2}$)) double ionisation transition. This is in contrast to the isovalent $\mathrm{SO_2^{2+}}$ dication for which the potential of the ground singlet state is free from any interaction with the upper states resulting in detectable long-lived $\mathrm{SO_2^{2+}}$ ions in our experiments. \cite{jarraya2021state,wallner2022abiotic} Quite different fragmentation mechanisms are expected for \ozonepp compared to $\mathrm{SO_2^{2+}}$.

\begin{table}[]
\centering
\caption{Vertical double ionization energies (VDIEs) from the 40.81 eV experimental spectrum adjacent to calculated ADIE and VDIEs of \ozonepp electronic states inside the 35 to 41 eV range of binding energy. }
\begin{threeparttable}[b]
\begin{tabular}{|l||l|ll|l|l|}
\hline
                      & Experiment \tnote{a)}                     & \multicolumn{2}{l|}{\centering Theory \tnote{a)} }                                           & Ref \cite{newson1996formation}              & Ref \cite{pyykko1989ab}      \\ \hline \hline
Electronic state      & VDIE (eV)                      & \multicolumn{1}{l|}{ADIE (eV) \tnote{b)}}             & VDIE (eV) \tnote{c)}              & ADIE (eV)         & ADIE (eV) \\ \hline
$\mathrm{X^1\Sigma_g^+}$                 &                                & \multicolumn{1}{l|}{34.445}                  & 35.893*                  & 34.3 ± 0.3        &           \\ \hline
\multirow{4}{*}{A3B2} & \multirow{4}{*}{35.5   ± 0.4} & \multicolumn{1}{l|}{\multirow{4}{*}{35.513}} & \multirow{4}{*}{35.603*} & \multirow{4}{*}{} & 35.17 \tnote{d)}   \\ \cline{6-6} 
                      &                                & \multicolumn{1}{l|}{}                        &                          &                   & 30.70 \tnote{e)}   \\ \cline{6-6} 
                      &                                & \multicolumn{1}{l|}{}                        &                          &                   & 35.6 \tnote{f)}    \\ \cline{6-6} 
                      &                                & \multicolumn{1}{l|}{}                        &                          &                   & 35.8 \tnote{g)}    \\ \hline
$\mathrm{1^3A_2}$                  & 36.4 ± 0.10                    & \multicolumn{1}{l|}{}                        & 36.713*                  &                   &           \\ \hline
$\mathrm{1^3B_1 }$                 &                                & \multicolumn{1}{l|}{}                        & 36.943                   &                   &           \\ \hline
$\mathrm{1^1A_2 }$                 &                                & \multicolumn{1}{l|}{}                        & 37.826*                  &                   &           \\ \hline
$\mathrm{2^3B_2 }$                 & 38 ± 0.02                      & \multicolumn{1}{l|}{}                        & 38.286                   &                   &           \\ \hline
$\mathrm{1^1B_2  }$                &                                & \multicolumn{1}{l|}{}                        & 38.429                   &                   &           \\ \hline
$\mathrm{2^1A_1 }$                 &                                & \multicolumn{1}{l|}{}                        & 38.449                   &                   &           \\ \hline
$\mathrm{1^1B_1}$                 &                                & \multicolumn{1}{l|}{}                        & 38.465                   &                   &           \\ \hline
$\mathrm{1^3A_1 }$                 &                                & \multicolumn{1}{l|}{}                        & 38.674*                  &                   &           \\ \hline
$\mathrm{2^3A_2 }$                 &                                & \multicolumn{1}{l|}{}                        & 39.113                   &                   &           \\ \hline
$\mathrm{3^1A_1}$                  &                                & \multicolumn{1}{l|}{}                        & 39.766*                  &                   &           \\ \hline
$\mathrm{2^1A_2  }$                &                                & \multicolumn{1}{l|}{}                        & 39.774                   &                   &           \\ \hline
$\mathrm{2^3B_1 }$                 &                                & \multicolumn{1}{l|}{}                        & 39.914                   &                   &           \\ \hline
$\mathrm{4^1A_1 }$                 &                                & \multicolumn{1}{l|}{}                        & 40.047                   &                   &           \\ \hline
$\mathrm{2^3A_1 }$                 &                                & \multicolumn{1}{l|}{}                        & 40.421                   &                   &           \\ \hline
$\mathrm{3^3B_1 }$                 &                                & \multicolumn{1}{l|}{}                        & 40.570                   &                   &           \\ \hline
$\mathrm{3^3A_1 }$                 &                                & \multicolumn{1}{l|}{}                        & 40.767                   &                   &           \\ \hline
$\mathrm{2^1B_1 }$                 &                                & \multicolumn{1}{l|}{}                        & 40.918*                  &                   &           \\ \hline
\end{tabular}
\begin{tablenotes}
\item [a)] This work
\item [b)] (R)CCSD(T)-F12/aVTZ (opt) // (R)CCSD(T)/aVTZ + $\Delta$CV + $\Delta$SR + $\Delta$ZVPE (SP) composite scheme
\item [c)] MRCI-F12/aug-cc-pVTZ calculations with (R)CCSD(T)-F12/aVTZ (opt) // (R)CCSD(T)/aVTZ + $\Delta$CV + $\Delta$SR composite scheme used as reference for VDIE of A3B2 state.
\item [d)] HF/6-31G*
\item [e)] MP2/6-31G*
\item [f)] MP3/6-31G*
\item [g)] CISD/6-31G*
\end{tablenotes}
  \end{threeparttable}
\label{tab:majdistable}
\end{table}

Moreover, Figure \ref{fig:angle_theory} shows that there are several crossings between the upper electronic states of \ozonepp occurring in the vicinity of the F.C. zone. Therefore, the experimental bands between 36 and 42 eV should arise from a dense manifold of strongly interacting states whose individual contributions cannot be disentangled from the present spectra alone. 

Figure \ref{fig:O3_mol} shows the optimised structures of the ground state of neutral ozone and those of the lowest singlet and triplet electronic states of ozone dication as computed at the (R)CCSD(T)-F12/aug-cc-pVTZ level. The total energies and geometries in cartesian coordinates at equilibrium are listed in Table S1 in Supplementary materials. This figure and this table show that the ground electronic state of \ozonepp is a linear singlet of $\mathrm{^1\Sigma_g^+}$ symmetry species, whereas the lowest triplet is bent and has a $\mathrm{^3B_2}$spectroscopic term. This is consistent with the earlier theoretical findings of Pyykkö et al. \cite{pyykko1989ab,pyykko1990ab} and of Champkin et al.\cite{champkin1999dissociation}. 

Figure \ref{fig:O3_mol} shows that the OO distances are slightly shortened upon double ionising \ozone($\mathrm{X^1A_1}$) by ~0.1 Å whereas the OOO in-plane angle is strongly changed from 117.2\textdegree to 180\textdegree. This results in unfavourable Franck-Condon factors (FCFs) upon single photon double ionising O$_3$($\mathrm{X^1A_1}$). Consequently, the 0-0 transition is most likely to be observable in the experimental spectra as discussed for the isovalent $\mathrm{SO_2}$/$\mathrm{SO_2^{2+}}$ system. \cite{hochlaf2004theoretical} In contrast, the first \ozonepp triplet has close equilibrium geometry as the neutral resulting in favourable FCFs. This is further confirmed upon computing the ADIEs to populate the \ozonepp($\mathrm{X^1\Sigma_g^+, A^3B_2}$) states. At the (R)CCSD(T)-F12b/aug-cc-pVTZ (opt) // (R)CCSD(T)/aug-cc-pVTZ + $\Delta$CV + $\Delta$SR + $\Delta$ZVPE (SP) level, these ADIEs are determined to be 34.445 and 35.513 eV, respectively \textcolor{blue}{(Table 2??ADIE)}. The corresponding VDIEs are 35.910 and 35.603 eV (cf. Table S2 in the Supplementary Materials). Table \ref{tab:majdistable} lists also the VDIEs of the upper electronic states of \ozonepp, which were obtained by adding the MRCI-F12/aug-cc-pVTZ relative energies on top of the VDIE of the \ozonepp $\mathrm{X^1\Sigma_g^+}$ state as evaluated using the (R)CCSD(T)-F12b/aug-cc-pVTZ (opt) // (R)CCSD(T)/aug-cc-pVTZ + $\Delta$CV + $\Delta$SR + $\Delta$ZVPE (SP) composite scheme. This table presents a high density of dicationic electronic states in the FC region leading most likely to congestion of the corresponding bands in the experimental spectra in Figure \ref{fig:ozoneDIE48eV} (* denotes the electronic states shown in Figure 3). 

\section{Electron-ion coincidences} 
\FloatBarrier

A mass spectrum obtained at the photon energy of 41 eV from either single or two ions detected in coincidence with two electrons can be found in the Supplementary Materials.
Signal from air is present in these data as reflected by the features at $\mathrm{m/q} = 18$ ($\mathrm{H_2O}$), 28 ($\mathrm{N_2}$), and 44 ($\mathrm{CO_2}$); the signal at $\mathrm{m/q} = 4$ arises from He leaking in from our gas-discharge lamp. 
Since no trace of $\mathrm{O_3^{2+}}$ (which should appear at $\mathrm{m/q} = 24$) is observed in either spectrum, these data illustrate that no dicationic electronic state of nascent $\mathrm{O_3^{2+}}$ is sufficiently long-lived to be detected as stable ion on the time scale of our mass spectrometer. This is in accord of the non-observation of \ozonepp formed by electron impact of \ozone in the experiments of Price and co-workers. \cite{newson1995electron} 

\begin{figure}[H]
    \centering
    \includegraphics[width=0.75\linewidth]{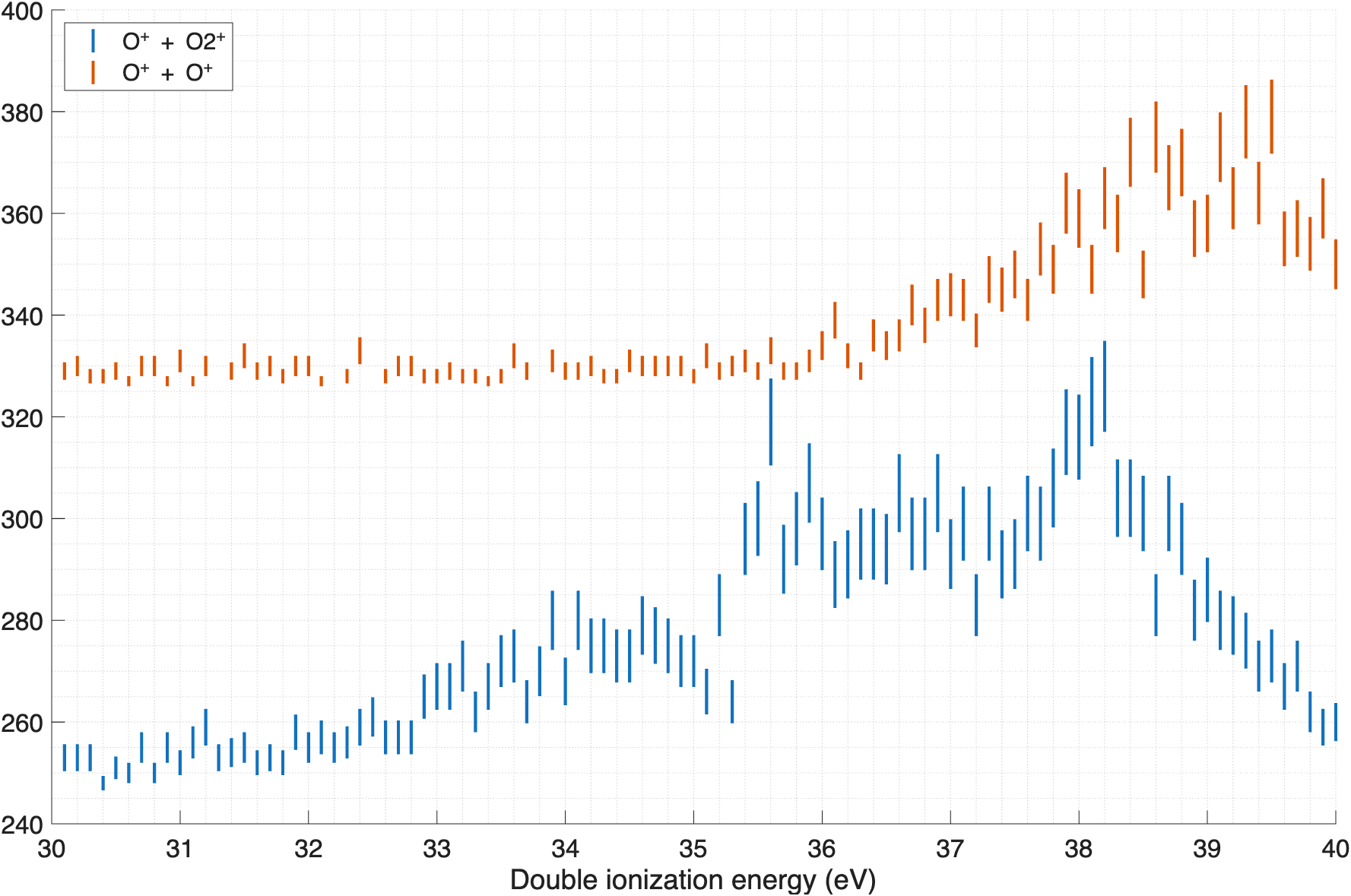}
    \caption{
    Double-ionization electron spectra from fourfold coincidence events. The blue spectrum corresponds to the upper-right selection in the ion-ion map, associated with $\mathrm{O^+ + O_2^+}$ ion coincidences. The orange spectrum shows the double-ionization energy for the lower-right selection, associated with $\mathrm{O^+ + O^+}$ ion coincidences. For the $\mathrm{O^+ + O_2^+}$ coincidences, a structure with peaks starting at 35.5~eV and a sharper peak near 38~eV is observed. In contrast, for the $\mathrm{O^+ + O^+}$ coincidences, the contribution to the total double-ionization electron signal begins at higher energy, around 36.4~eV, and peaks above 38.5~eV. The wings of the electrons in coincidence with the wings (?) were used for the double ionization energy calculation, to avoid contribution from background that was visible in the full ion island.}
    \label{fig:frommapselection}
\end{figure}

Figure \ref{fig:frommapselection} presents the results of double-ionization measurements on ozone obtained in electron-ion coincidence mode. The analysis is based on fourfold coincidence events, enabling a direct correlation between the double-ionization energy and specific ion-ion fragmentation channels.

The time-of-flight ion-ion coincidence maps from this data selection, which reveal the different dissociation pathways, can also be found in the Supplementary Materials.
 The left panel corresponds to $\mathrm{O^+ + O_2^+}$ coincidences, while the right panel shows $\mathrm{O^+ + O^+}$ coincidences. In the right panel, only the outer regions of the kinetic-energy-release islands are selected, emphasizing contributions from higher-energy dissociation pathways. The slopes of the red line in these maps are $-1$, indicating that the dissociation predominantly proceeds via a direct breakup channel.

The corresponding double-ionization electron spectra, selected based on these ion-ion correlations, are shown in Fig.~\ref{fig:frommapselection}. For the $\mathrm{O^+ + O_2^+}$ channel, the spectra exhibit a structured onset at 35.5~eV, followed by a pronounced peak near 38~eV. The signal below 35 eV stems from a two step process: $\mathrm{O_3 \rightarrow O_3^{+*} \rightarrow O^* + O_2^+: O^* \rightarrow O+}$, as supported by the electron data where no energy sharing is detected below 35 eV. This process has been identified in measurements on $\mathrm{O_2}$ and is thoroughly explained by Feifel et al. \cite{feifel2005valence} This is most probably more pronounced here than in the electron only measurements because of the slighly higher pressures used for these measurements, as mentioned earlier. In contrast, the $\mathrm{O^+ + O^+}$ channel contributes at higher energies, with an onset around 36.4~eV and a maximum above 38.5~eV. These results demonstrate that the double-ionization dynamics depend sensitively on the selected ion-ion channel, highlighting the power of multi-particle coincidence measurements for disentangling complex dissociation processes.

As shown, when selecting $\mathrm{O^+ + O_2^+}$ coincidences, the features associated with electron-only double ionization are clearly reproduced, most prominently in the blue spectrum. In contrast, for the $\mathrm{O^+ + O^+}$ channel, spectral features do not emerge until above 36.4~eV, with the main contribution appearing only beyond the sharp peak near 38~eV. This behavior is unexpected, given that the thermodynamic thresholds for these processes are significantly lower as given in Table~\ref{tab:thermo}. For the dissociation limits, the true 0 K values were deduced similarly to Berkowitz et al. \cite{berkowitz2008absolute} by using the dissociation energy of $\mathrm{O_3 \rightarrow O(^3P_2) + O_2(X ^2\Pi_g^-)}$ of 1.0621(4) eV \cite{taniguchi1999determination} as the starting point for all four processes listed in Table \ref{tab:thermo}. For $\mathrm{O_3 \rightarrow O_2^+ + O^+}$, the first ionization energy of $\mathrm{O_2\,(X^2\Pi_g)}$ of 12.07014(15) eV \cite{merkt1998high} and the first ionization energy of $\mathrm{O\,(^4S)}$ of 13.618055(7) eV \cite{MooreCharlotteEmma18981958Aela} are used to produce a 0 K appearance energy (AE) of 26.75 eV. In the case of $\mathrm{O_3 \rightarrow O + 2O^+}$, the first dissociation limit of $\mathrm{O_2^+ \rightarrow O(^3P) + O^+(^4S^o)}$ of 18.734 eV \cite{blyth1981competing} replaces the use of the first ionization energy of $\mathrm{O_2}$, leading to a 0 K dissociation limit of 33.41 eV. The two last processes in Table \ref{tab:thermo} simply utilize the neutral dissociation energy and the first double ionization energies of the resultants.  $\mathrm{O_2}$ has a first double ionization energy of 36.13 $\pm$ 0.02 eV \cite{blyth1981competing} resulting in a 0 K dissociation limit of 37.19 eV for $\mathrm{O_3 \rightarrow O_2^{2+} + O}$ and the second ionization energy of $\mathrm{O}$ is reported as  283 270.9 $\pm$ 0.5 cm$^{-1}$ (35.1211 eV)\cite{wenaaker1990spectrum}, giving a 0 K dissociation limit of 49.8 eV for $\mathrm{O_3 \rightarrow O^{2+} + O_2}$. The corresponding computed values at the (R)CCSD(T)-F12/aug-cc-pVTZ level of theory compare favorably to those described in this section. Indeed, the differences are less than 0.1 eV between both data sets.

\begin{table}[h]
    \centering
    \caption{Comparison of the calculated thermodynamic thresholds (True 0K dissociation energies) for the processes identified in the experimental data \textcolor{blue}{(, together with the measured appearance energies (AE) )}and their corresponding associated kinetic energy releases (KER) as derived from the peak widths in the ion-ion coincidence maps. The first reaction has been split into two different DIE regions, 34-36~eV (*) and 37 to 40~eV (**), as indicated by the asterisks. 
    The True 0K dissociation energies makes use of identified appearance energies noted in the literature together with the dissociation energy associated with the $\mathrm{O_3 \rightarrow O + O_2}$ process. \cite{berkowitz2008absolute,MooreCharlotteEmma18981958Aela,blyth1981competing,merkt1998high,hall1992vibrational,wenaaker1990spectrum,taniguchi1999determination} The calculated corresponding data are found through our (R)CCSD(T)-F12/aug-cc-pVTZ calculations where the identified asymptotes have been listed. The third column lists the KER, derived from the peak widths in the ion-ion coincidence maps. }

    \begin{threeparttable}[b]
\begin{tabular}{|l|l|l|l|l|}
\hline
\multirow{2}{*}{\textbf{Process}}        & \textbf{True 0K dissociation}  & \textbf{Calculated dissociation}  &  \multirow{2}{*}{\textbf{KER} \tnote{b)}  }        & \multirow{2}{*}{\textbf{Literature KER (eV) }\tnote{c)}} \\
                                &        \textbf{energy (eV)} \cite{berkowitz2008absolute,MooreCharlotteEmma18981958Aela,blyth1981competing,merkt1998high,hall1992vibrational,wenaaker1990spectrum}  & \textbf{ energy (eV)} \tnote{a)}    &              &                         \\
                                \hline
\(\mathrm{O_3 \rightarrow O_2^+ + O^+}\) (*)           & 26.75                                              & 26.66                                              & 6.0                                      & \multirow{2}{*}{\shortstack{7.5±0.3 \\  7.5±0.3 \\ 8.1±0.4}} \\[1.2 ex]
\cline{1-4}
\(\mathrm{O_3 \rightarrow O_2^+ + O^+}\) (**)        & 26.75                                              & 26.66                                              & 6.8                                      &                          \\[1.2ex]
\hline 
\(\mathrm{O_3 \rightarrow 2O^+ + O}\)             & 33.41                                              & 33.33                                              & 2.8                                      &                          \\
\hline
\(\mathrm{O_3 \rightarrow O_2^{2+} + O}\)           & 37.19                                              & 37.12                                              & -                                        &                          \\
\hline
\(\mathrm{O_3 \rightarrow O^{2+} + O_2}\)            & 49.80                                              & 49.72                                              & -                                        &                         \\
\hline

\end{tabular}
\begin{tablenotes}
\item [a)] Calculations from this work
\item [b)] Experimental data from this work
\item [c)] Electron impact at 55, 150 and 233 eV of electron energy \cite{champkin1999dissociation}
\end{tablenotes}
\end{threeparttable}

        \label{tab:thermo}
\end{table}

The spectra in Figure \ref{fig:frommapselection} may be explained by the kinetic energy release (KER) associated with the respective processes, as inferred from the peak widths. The KER for the $\mathrm{O_3 \rightarrow O_2^+ + O^+}$ channel has been split up into a lower and higher double ionization energy (DIE) region to evaluate the KER in the first and the last feature of the spectrum in Figure \ref{fig:frommapselection}. The lower DIE region, 34-36 eV, is estimated to be 6.0~eV, and the higher DIE region, 37-40 eV, has an ion peak width correlating to 6.75 eV KER. The KER for the  $\mathrm{O_3 \rightarrow 2O^+ + O}$  process has been deducted as a whole and is approximately 2.8~eV. The slope of the ion–ion coincidence islands is $-1$, indicating that both processes proceed via two-body dissociation.\cite{olsson2024anisotropy}  In the case of the $\mathrm{O_3 \rightarrow 2O^+ + O}$ channel, this implies either that the neutral O atom departs prior to the dissociation of the transient $\mathrm{O_2^{2+}}$ dication, or that the neutral O atom is a pure spectator in immediate dissociation. It might be the central atom of linear \ozonepp. The KER values are estimated from the full width at half maximum (FWHM) of the ion–ion coincidence islands, expressed as t2 – t1 (PIPICO peak widths), where the thermal translational velocity contribution cancels. This method of deducing KER is known to risk underestimating the energy released since the high energy wings may be overlooked, so each KER in this article serves as a lower bound. When the KER is added to the corresponding 0 K thresholds, this gives an indication to what the internal energy of the outgoing ions from the reaction may be. We note that the KER measured in electron impact measurements by Newson and Price\cite{newson1996formation} are slightly higher than what we have derived from our measurements. They have measured the KER in the same way that we have, but the ionization mechanism is different. This could be the cause of the discrepancy in ionization method, as this is also the case for the double ionization energy found in the same paper at 34.3 eV compared to both our first vertical double ionization value at 35.5 eV and appearance energy at about 35.1 eV.\cite{newson1996formation} 

\begin{figure}[H]
    \centering
    \includegraphics[width=1\linewidth]{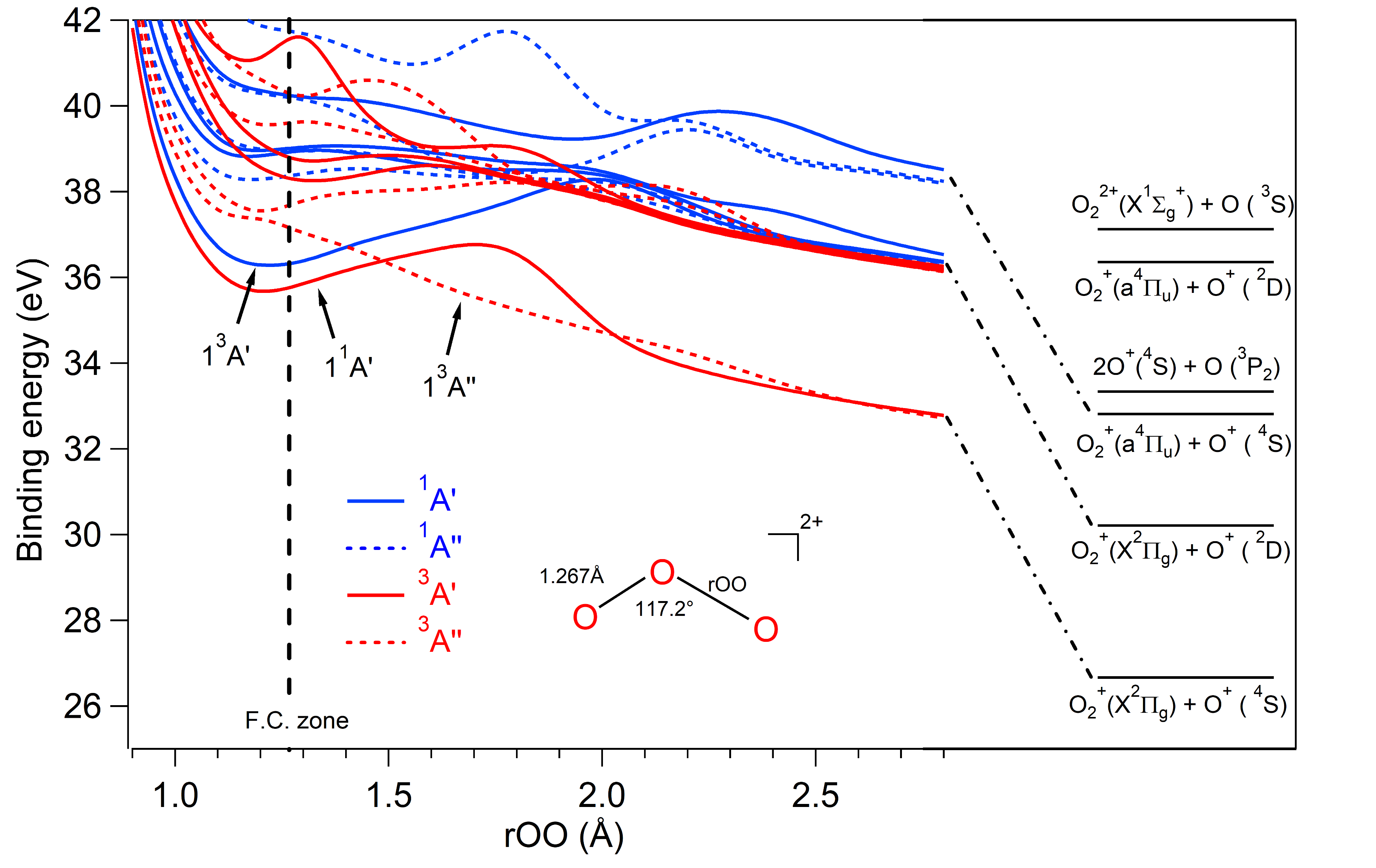}
    \caption{1D PES cut of the potential of the O$_3^{2+}$ electronic excited states obtained at the MRCI/aug-cc-pV5Z level of theory along the stretching of one internal O-O bond while the other stays fixed at its value obtained for the geometry optimization of O$_3$ (X$^1A_1$) at the CCSD(T)-F12/aug-cc-pVTZ level of theory.}
    \label{fig:rOO_theory}
\end{figure}

Using the data in Table \ref{tab:thermo} together with the potential energy surfaces (PES) of $\mathrm{O_2^+}$\cite{GILMORE1965369} and the known energies of ground and excited states of $\mathrm{O^+}$\cite{MooreCharlotteEmma18981958Aela}, some conclusions can be drawn on the final ion states in the dissociation processes that have been identified. In addition, calculations have been performed that are presented in Figure \ref{fig:rOO_theory} where the one-dimensional cuts of the 3D potential energy surfaces of \ozonepp dication along the OO distance is presented, whilst the other OO bond and the OOO in-plane angle are kept fixed at 1.267 Å and 117.2\textdegree \, i.e., their values in \ozone ($\mathrm{X^1A_1}$) at equilibrium. The figure shows a high density of electronic states is noticeable leading to their mutual interactions and the mixing of their electronic wavefunctions. For instance, the first $\mathrm{^3A''}$ state is repulsive and crosses the lowest singlet and lowest triplet not far in the FC region, leading to the predissociation. This prevents detecting long-lived \ozonepp in the present experiments and those of Newson and Price \cite{newson1996formation} as pointed out by Champkin et al. \cite{champkin1999dissociation}.

Looking at the $\mathrm{O_3 \rightarrow O_2^+ + O^+}$ process, the first part of the spectrum in the 34-36 eV DIE region would leave 2.75 eV available after the dissociation for internal energy, see Table \ref{tab:thermo}. If the process isn't adiabatic, and the bond length in ozone extends to at least 1.5 Å before the dissociation, the ground state in $\mathrm{O_2^+}$ can hold the excess energy, accounting for the entirety of the internal energy. This is also consistent with the (PES) in Figure \ref{fig:rOO_theory}  where the lowest state, $\mathrm{1^1A'}$ as well as the repulsive state $\mathrm{1\;^3A''}$, dissociates to $\mathrm{O_2^+}$ and $\mathrm{O^+}$ in the ground state with an elongation of the O-O bond to 1.8 Å. This elongation would enable up to 4.5 eV as vibrational energy in $\mathrm{O_2^+\;X^2\Pi_g}$. In addition to the vibrational energy, the $\mathrm{O_2^+}$ ion can also hold some rotational energy which is probable if $\mathrm{O_3^{2+}}$ is not linear upon dissociation. 

Upon formation of \ozonepp ($\mathrm{A^3B_2}$ (1 $\mathrm{^3A’\; in\; C_s}$ point group)), the reaction may follow the potential of this state and leads to  $\mathrm{O_2^+ \;(X^2\Pi_g) + O^+ \; (^4S)}$ after overcoming the potential barrier of the ($\mathrm{A^3B_2}$ (1 $\mathrm{^3A’}$) state. Instead, the \ozonepp (($\mathrm{A^3B_2}$ (1 $\mathrm{^3A’}$)) may be converted into \ozonepp (1 $\mathrm{^3A’'}$) at their respective crossing and then the reaction takes place on the repulsive \ozonepp (1 $\mathrm{^3A’'}$) potential. Such conversion is allowed by spin-orbit (cf. Table \ref{tab:so} in Supplementary materials). This lowers the KER and the AE by few tenths of eV. The KERs associated with both pathways remain far from the measured ones (by > 3.5 eV, Table \ref{tab:calcker}. Nevertheless, close agreement between the computed and measured KERs (~6.55 or 6.16 eV vs. $\sim$ 6 eV) if we consider the production of $\mathrm{O_2^+ \;(X^2\Pi_g) + O^+ \; (^2D)}$. Alternatively, the  $\mathrm{ O_2^+ \;(X^2\Pi_g)}$ ions may be formed in high vibrational levels carrying thus > 3.5 eV internal energy. Upon formation of \ozonepp $X^2\Sigma_g^+$ (1 $^1A’$)), we propose that these ions are converted by spin-orbit into the \ozonepp $\mathrm{A^3B_2}$ (1 $\mathrm{^3A’}$)) at the singlet – triplet crossing and then the reaction takes place as describe just above. Table \ref{tab:calcker} shows that such conversion is possible and that a KER of 6.50 is associated with such mechanism. According to the PESs in Figure \ref{fig:rOO_theory}, 1 $\mathrm{^1A’}$ exhibits a potential well of ~1.1 eV depth. Whereas, we compute a much deeper potential (of ~2 eV) for the 1 $\mathrm{^3A’}$ \textcolor{purple}{1 $\mathrm{^1A’}$ (= $\mathrm{X^1\Sigma_g^+}$)} state, which correlates at large OO distances to the excited $\mathrm{O_2^+ \;(X ^2\Pi_g) + O^+ \; (^2D)}$ asymptote. Assuming that the fragmentation occurs on the singlet (triplet) potential, an AE and a KER of ~38.3 eV (~36.7 eV) and of ~8.1 eV (10.1 eV) can be deduced from these potentials are (cf. Table \ref{tab:calcker}). Both sets of values are different from those measured previously and in the present work. Obviously, multi-step mechanisms, where several electronic states are involved, should be invoked instead. 

The latter part of the spectrum, from the 37-40 eV DIE region, leaves 4.5 eV as internal energy after the dissociation, see Table \ref{tab:thermo}. Based on the F.C. of ozone combined with the PES of $\mathrm{O_2^+}$ and the energies of $\mathrm{O^+}$, the resultant ions appear to be $\mathrm{O_2^+ \;(X ^2\Pi_g) + O^+ \; (^2D)}$ with 1.2 ($\mathrm{\nu = 5}$) and 3.33 eV internal energy, respectively, also reflected in the PES of $\mathrm{1^3A'}$ in Figure \ref{fig:rOO_theory}. Based on the PES in Figure \ref{fig:rOO_theory}, some dissociations above 37 eV also result in $\mathrm{O_2^+ \;(a^4\Pi_u) + O^+ \; (^4S)}$, where the elongation of the O-O bond would allow for an internal energy of up to 2 eV, easily accounting for the excess of 4.5 eV after the dissociation.  However, the high density of states above 37 eV makes it complicated to point to a single undertaken mechanism in the dissociation. Examination of Table \ref{tab:calcker} in Supplementary materials reveals that a KER of 6.63 eV (close to the measured 6.8 eV) when the reaction occurs on the \ozonepp ($3^1A''$) potential leading to $\mathrm{O_2^+ \;(X^4\Pi_u) + O^+ \; (^4S)}$ excited asymptote. The the \ozonepp ($3^1A''$) state may be populated either directly upon doubly ionizing \ozone \, or after internal conversion from the upper \ozonepp states. 

The $\mathrm{O^+ + O^+}$ reaction leaves 1.8 eV in internal energy after the dissociation, since none of the states agree with this kind of internal energy, the remaining O atom is assumed to have kept some of the energy in the dissociation, leaving both the ions in ground state, $\mathrm{O^+\, (^4S)}$.

\begin{figure}[H]
    \centering
    \includegraphics[width=1\linewidth]{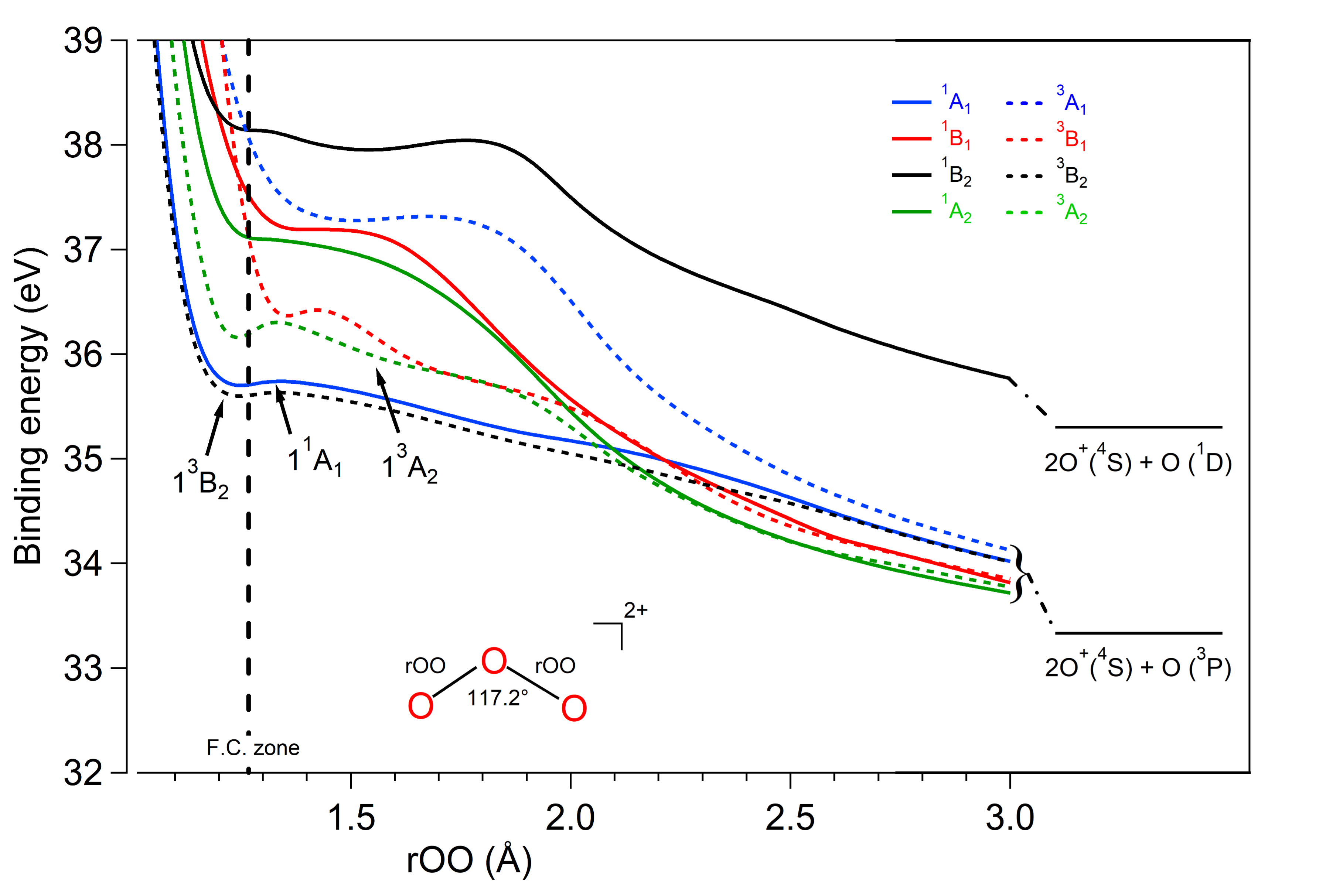}
    \caption{CASSCF/aug-cc-pVTZ one dimensional cuts of the potential energy surfaces of the lowest electronic states of \ozonepp while lengthening simultaneously both OO bond. The OOO in-plane angle is kept fixed at its equilibrium value in \ozone ($X^1A_1$). The reference energy is that of \ozone ($X^1A_1$) at equilibrium.}
    \label{fig:rOO_O_theory}
\end{figure}

Figure \ref{fig:rOO_O_theory} shows the one-dimensional evolution of the singlet and triplet states of the \ozone \, dication while lengthening both OO distances symmetrically. These potentials correlate at large interatomic distances to the $\mathrm{O^+ + O^+ + O}$ asymptote. This figure shows that upon population of the dicationic potential wells in the F.C. zone, small barriers separate them for the 1/R Coulombic evolution before reaching the respective dissociation limit. In particular, the \ozonepp electronic state lying in the 35.5 – 38 eV range lead to the $\mathrm{O^+ + O^+ + O}$ fragments in their electronic ground state. Whereas the lowest $\mathrm{^1B_2}$ state lead adiabatically to the $\mathrm{O^+ + O^+ + O(^1D)}$ excited channel. Table \ref{tab:calcker} lists the calculated KERs and AEs deduced from these potentials for the $\mathrm{X^1\Sigma_g^+, A^3B_2 \; and\; 3^1A''}$ states i.e., KER = 2.41, 2.30 and 2.97 eV, respectively, in agreement with the 2.8 eV determined in the present experiments.

\section*{Conclusions} 

In this work, we have substantially extended the experimental and theoretical characterization of ozone into the regime of valence double photoionization. By combining HeII-$\alpha$, HeII-$\beta$, and higher-energy vacuum ultraviolet radiation with a versatile multiple charged-particle correlation detection technique, we have recorded the first single-photon valence double ionization electron spectrum of O$_3$, and obtained information on the fragmentation behavior of the dicationic states. The measured double ionization energies provide new experimental benchmarks for theoretical treatments of multiply ionized ozone.

To interpret the experimental observations, we mapped the lowest potential energy surfaces of O$_3^{2+}$ using post-Hartree-Fock multi-configurational interaction approaches and determined with high accuracy the energetics of the relevant dissociation channels. The combined experimental and theoretical results reveal that dissociative double ionization of ozone proceeds not only through the formation of ground-state O$_2^+$ + O$^+$ fragments, but also through channels producing electronically excited O$^+$ ions. These findings demonstrate that the fragmentation dynamics of doubly ionized ozone are considerably richer than previously recognized.

The present results are expected to contribute to a deeper understanding of the energy redistribution and decay pathways following double ionization of ozone, with direct implications for ion-molecule chemistry in the Earth’s ionosphere and in extraterrestrial planetary atmospheres exposed to intense ionizing radiation fields. In particular, the observation of electronically excited atomic oxygen fragments may be of broader atmospheric relevance, as excited oxygen species are known to play a central role in driving highly reactive chemical pathways. In this context, the present findings may bear analogy to the well-known photodissociation of neutral O$_3$ producing O($^1$D) atoms, which constitute an important source of chemical activation in atmospheric chemistry.

Overall, the present study provides a basis for future experimental and theoretical investigations of reactive molecular dications in atmospheric and astrochemical environments.

\section*{Acknowledgements}
This work has been financially supported by the Swedish Research Council (grant number 2023-03464), the Carl Tryggers Foundation (grant number 24:3719), the Adlerbertska Foundation (grant holder Richard Squibb), and the Knut and Alice Wallenberg Foundation (grant numbers 2017.0104 and 2024.0120), Sweden. The data handling of the quantum chemical calculations was enabled by resources provided by the National Academic Infrastructure for Supercomputing in Sweden (NAISS), partially funded by the Swedish Research Council through grant agreement no. 2022-06725. The computational resources provided by C3SE are also acknowledged. We thank the Ongoing Research Funding program, (ORF-2026-808), King Saud University, Riyadh, Saudi Arabia.

\bibliography{sample}

\newpage
\appendix

\Large{The supplementary materials.}

\begin{figure}[H]
    \centering
    \includegraphics[width=0.75\linewidth]{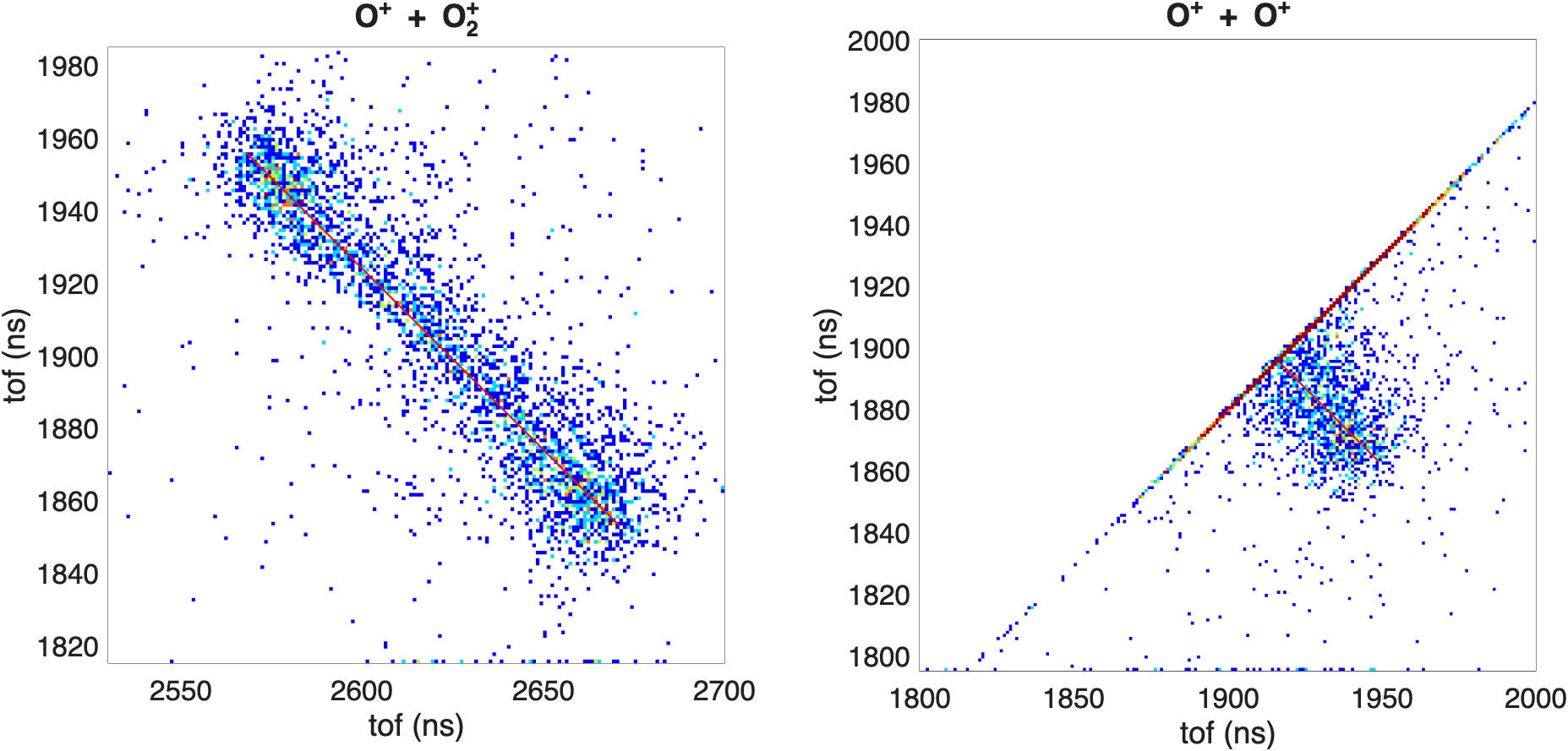}
    \caption{Time-of-flight ion-ion coincidence maps obtained from fourfold electron-ion events. The left
    panel corresponds to $\mathrm{O^+ + O_2^+}$ coincidences, while the right panel shows $\mathrm{O^+ + O^+}$ coincidences. The red lines have a slope of $-1$, indicating a predominantly direct dissociation process.
    }
    \label{fig:ionmaps}
\end{figure}

\begin{figure}[H]
    \centering
    \includegraphics[width=0.75\linewidth]{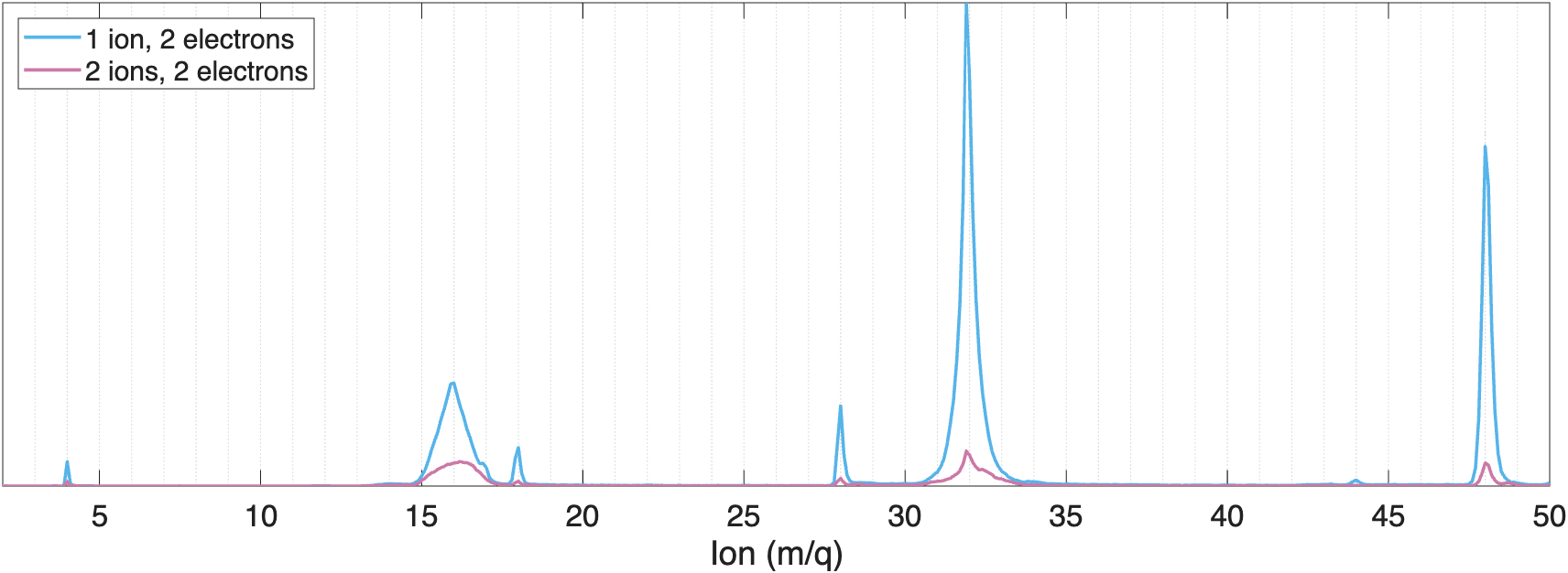}
    \caption{Mass spectrum of both single and two ions detected in coincidence with two electrons. Signals from air are visible in the single-ion–two-electron spectrum at $m/q = 18$ ($\mathrm{H_2O}$), 28 ($\mathrm{N_2}$), and 44 ($\mathrm{CO_2}$); the signal at $m/q = 4$ arises from He in the gas-discharge lamp. No trace of $\mathrm{O_3^{2+}}$ (which would appear at $m/q = 24$) is observed in either spectrum. }
    \label{fig:ionspectrum}
\end{figure}

Result (click "Generate" to refresh)
Copy to clipboard

\begin{table}[]
\centering
\caption{Cartesian coordinates matrix obtained after geometry optimisation at the (R)CCSD(T)-F12b/aug-cc-pVTZ level of theory. Absolute energy in Hartree and harmonic frequencies with their respective symmetry are also given.}
\begin{tabular}{|l|l l l l|}
\hline
\multirow{6}{*}{\ozone ($\mathrm{X^1A_1}$)}    & \multicolumn{4}{c|}{-225.21594662 }                                                  \\ \cline{2-5} 
                              & O     &     0.0000000000    &   -0.0000000000    &    0.4400339364 \\ \cline{2-5} 
                              & O     &     0.0000000000     &   1.0810164025    &   -0.2200169682 \\ \cline{2-5} 
                              & O    &      0.0000000000     &  -1.0810164025    &   -0.2200169682 \\ \cline{2-5} 
                              & & $\mathrm{\nu_1 \;(a_1)}$      &      $\mathrm{\nu_2 \;(a_1)}$    &     $\mathrm{\nu_3 \;(b_2)}$                \\ \cline{2-5} 
                              & &1175.53   &   727.92    &   1095.11                               \\ \hline
\multirow{6}{*}{\ozonepp ($\mathrm{X^1\Sigma_g^+}$)} & \multicolumn{4}{c|}{-223.95164348   }                                                \\ \cline{2-5}
                              & O    &      0.0000000000  &    -0.0000000000  &      0.0000654224 \\ \cline{2-5}
                              & O    &      0.0000000000  &      1.1676220132 &      -0.0000327112 \\ \cline{2-5} 
                              & O    &     0.0000000000  &     -1.1676220132 &      -0.0000327112 \\ \cline{2-5} 
                              & &$\mathrm{\nu_1 \;(\sigma_g^+)}$  &     $\mathrm{\nu_2 \;(\pi_u)}$     &   $\mathrm{\nu_3 \;(\sigma_u^+)}$                          \\ \cline{2-5}
                              & &1056.73    &  417.29    & 1654.55                                 \\ \hline
\multirow{6}{*}{\ozonepp ($\mathrm{A^3B_2}$)}  & \multicolumn{4}{c|}{-223.90957446  }                                                 \\ \cline{2-5} 
                              & O    &      0.0000000000  &      0.0000000000  &      0.4172333388 \\ \cline{2-5} 
                              & O    &      0.0000000000  &      1.1086297756  &     -0.2086166694 \\ \cline{2-5}
                              & O    &      0.0000000000  &     -1.1086297756  &     -0.2086166694 \\ \cline{2-5} 
                              & &$\mathrm{\nu_1 \;(a_1)}$  &         $\mathrm{\nu_2 \;(a_1)}$  &        $\mathrm{\nu_3 \;(b_2)}$                     \\ \cline{2-5} 
                              & &982.40      & 559.40   &     586.47                               \\ \hline
\end{tabular}
\end{table}

\begin{table}[]
\caption{Adiabatic double ionization energy (ADIE in eV) of \ozonepp ($\mathrm{X^1\Sigma_g^+}$) and \ozonepp ($\mathrm{A^3B_2}$) calculated using the RCCSD(T)-F12/aug-cc-pVTZ (opt) // RCCSD(T)/ aug-cc-pVTZ + $\mathrm{\Delta}$CV + $\mathrm{\Delta}$SR + $\mathrm{\Delta}$ZVPE (SP) composite scheme. The vertical double ionization energy (VDIE in eV) of \ozonepp ($\mathrm{X^1\Sigma_g^+}$) and \ozonepp ($\mathrm{A^3B_2}$) are also calculated following the same composite scheme excluding the $\mathrm{\Delta}$ZVPE contribution}
\begin{tabular}{|l|ll|l|ll|}
\hline
                                                                                   & \multicolumn{1}{l|}{\ozonepp ($\mathrm{X^1\Sigma_g^+}$)} & \ozonepp ($\mathrm{A^3B_2}$)  & O3 ($\mathrm{X^1A_1}$)    & \multicolumn{1}{l|}{\ozonepp ($\mathrm{X^1\Sigma_g^+}$)} & \ozonepp ($\mathrm{A^3B_2}$)  \\ \hline
PBE0/aug-cc-pVTZ (opt)                                                             & \multicolumn{1}{l|}{-223.9956634} & -223.9741746 & -225.2714113 & \multicolumn{1}{l|}{}             &              \\ \hline
ZPE (PBE0/aug-cc-pVTZ) (eV)                                                        & \multicolumn{1}{l|}{0.226961179}  & 0.136104972  & 0.208847538  & \multicolumn{1}{l|}{}             &              \\ \hline
RCCSD(T)-F12/aug-cc-pVTZ (opt)                                                     & \multicolumn{1}{l|}{-223.9516435} & -223.9095745 & -225.2159466 & \multicolumn{1}{l|}{-223.8990107} & -223.9088958 \\ \hline
RCCSD(T,fc)/cc-pwCVTZ (SP)                                                         & \multicolumn{1}{l|}{-223.8983826} & -223.8577399 & -225.1481896 & \multicolumn{1}{l|}{-223.846658}  & -223.8566191 \\ \hline
RCCSD(T,full)/cc-pwCVTZ (SP)                                                       & \multicolumn{1}{l|}{-224.0550069} & -224.0138879 & -225.3059049 & \multicolumn{1}{l|}{-224.0014558} & -224.0128177 \\ \hline
RCCSD(T)/cc-pVTZ (SP)                                                              & \multicolumn{1}{l|}{-223.8836426} & -223.8431194 & -225.1326419 & \multicolumn{1}{l|}{-223.832442}  & -223.8419878 \\ \hline
RCCSD(T)/cc-pVTZ-DK (SP)                                                           & \multicolumn{1}{l|}{-223.8701241} & -223.8295547 & -225.1189072 & \multicolumn{1}{l|}{-223.8189043} & -223.8284305 \\ \hline
                                                                                   & \multicolumn{1}{l|}{}             &              &              & \multicolumn{1}{l|}{}             &              \\ \hline
                                                                                   & \multicolumn{2}{c|}{ADIE (eV)}                   &              & \multicolumn{2}{c|}{VDIE (eV)}                   \\ \hline
RCCSD(T)-F12/aug-cc-pVTZ (opt)                                                     & \multicolumn{1}{l|}{34.403}       & 35.548       &              & \multicolumn{1}{l|}{35.836}       & 35.567       \\ \hline
 $\mathrm{\Delta}$CV                                                                                & \multicolumn{1}{l|}{0.029687365}  & 0.042648427  &              & \multicolumn{1}{l|}{0.079389532}  & 0.041272075  \\ \hline
$\mathrm{\Delta}$SR                                                                                & \multicolumn{1}{l|}{-0.005883377} & -0.004627571 &              & \multicolumn{1}{l|}{-0.00536119}  & -0.004828391 \\ \hline
$\mathrm{\Delta}$ZVPE                                                                              & \multicolumn{1}{l|}{0.018113641}  & -0.072742566 &              & \multicolumn{1}{l|}{}             &              \\ \hline
\multirow{2}{*}{\shortstack{RCCSD(T)-F12/aug-cc-pVTZ (opt) //\\ RCCSD(T)/   aug-cc-pVTZ + $\mathrm{\Delta}$CV \\ + $\mathrm{\Delta}$SR + $\mathrm{\Delta}$ZVPE (SP)}} & \multicolumn{1}{l|}{34.445}       & 35.513       &              & \multicolumn{1}{l|}{35.910}       & 35.603       \\
 &  \multicolumn{1}{l|}{}     &       &              &    \multicolumn{1}{l|}{}   &       \\
  &  \multicolumn{1}{l|}{}     &       &              &  \multicolumn{1}{l|}{}     &       \\
 \hline
\end{tabular}
\end{table}

\begin{table}[]
\caption{Calculated KERs as deduced from the MRCI/aug-cc-pV5Z 1D PES cuts of ozone dication electronic states along the O-O bond (Figure 5), toward the different considered asymptote of $\mathrm{O_2^+ + O^+}$ fragmentation channel, from several starting points on the curves. }
\begin{threeparttable}[b]
\begin{tabular}{|l|lll|ll|}
\hline
                           & \multicolumn{3}{l|}{Starting from the top of the barrier of}                   & \multicolumn{2}{l|}{Starting from crossing between}            \\ \hline
Channel                    & \multicolumn{1}{l|}{$\mathrm{X^1\Sigma_g^+}$}      & \multicolumn{1}{l|}{$\mathrm{A^3B_2}$}       & $\mathrm{3^1A''}$      & \multicolumn{1}{l|}{$\mathrm{X^1\Sigma_g^+}$ - ($\mathrm{A^3B_2}$/$\mathrm{1^3A'}$)’} & $\mathrm{A^3B_2}$ - ($\mathrm{A^3B_2}$/$\mathrm{1^3A'}$) \\ \hline
$\mathrm{O_2^+}$ ($\mathrm{X^2\Pi_g}$) + $\mathrm{O^{+}}$ ($\mathrm{^4S}$)       & \multicolumn{1}{l|}{11.62   \tnote{a)}} & \multicolumn{1}{l|}{10.10   \tnote{b)}} & 12.78   \tnote{c)} & \multicolumn{1}{l|}{10.05   \tnote{d)}}           & 9.70   \tnote{e)}          \\ \hline
$\mathrm{O_2^+}$ ($\mathrm{X^2\Pi_g}$) + $\mathrm{O^{+}}$ ($\mathrm{^2D}$)       & \multicolumn{1}{l|}{8.08   \tnote{a)}}  & \multicolumn{1}{l|}{6.55   \tnote{b)}}  & 9.24   \tnote{c)}  & \multicolumn{1}{l|}{6.50   \tnote{d)}}            & 6.16   \tnote{e)}          \\ \hline
$\mathrm{O_2^+}$ ($\mathrm{X^4\Pi_u}$) + $\mathrm{O^{+}}$ ($\mathrm{^4S}$)       & \multicolumn{1}{l|}{5.47   \tnote{a)}}  & \multicolumn{1}{l|}{3.95   \tnote{b)}}  & 6.63   \tnote{c)}  & \multicolumn{1}{l|}{3.90   \tnote{d)}}            & 3.56   \tnote{e)}          \\ \hline
$\mathrm{O_2^+}$ ($\mathrm{X^4\Pi_u}$) + $\mathrm{O^{+}}$ ($\mathrm{^2D}$)       & \multicolumn{1}{l|}{1.92   \tnote{a)}}  & \multicolumn{1}{l|}{}           & 3.08   \tnote{c)}  & \multicolumn{1}{l|}{}                     &                    \\ \hline
$\mathrm{O_2^{2+}}$ (X1Sg+) + O ($\mathrm{^3S}$)      & \multicolumn{1}{l|}{}           & \multicolumn{1}{l|}{}           & 2.31   \tnote{c)}  & \multicolumn{1}{l|}{}                     &                    \\ \hline
$\mathrm{O^{+}}$ ($\mathrm{^4S}$) + $\mathrm{O^{+}}$ ($\mathrm{^4S}$) + O ($\mathrm{^3P}$) & \multicolumn{1}{l|}{2.41 \tnote{f,g)}}  & \multicolumn{1}{l|}{2.30 \tnote{f,h)}}  & 2.97 \tnote{f,i)} & \multicolumn{1}{l|}{}                     &                    \\ \hline
\end{tabular}
\begin{tablenotes}
\item [a)] Appearance energy = 38.28 eV
\item [b)] Appearance energy = 36.76 eV
\item [c)] Appearance energy = 39.44 eV
\item [d)] Appearance energy = 36.71 eV
\item [e)] Appearance energy = 36.40 eV
\item [f)] CASSCF/aug-cc-pVTZ level
\item [g)] Appearance energy = 35.74 eV
\item [h)] Appearance energy = 35.63 eV
\item [i)] Appearance energy = 36.30 eV
\end{tablenotes}
\end{threeparttable}
\label{tab:calcker}
\end{table}

\begin{table}[]
    \centering
    \caption{Spin-Orbit coupling matrix element between $\mathrm{X^1\Sigma_g^+ - 1^3A''}$ and $\mathrm{A^3B_2} - 1^3A''$ states at the position of their respective crossings. The other OO distance and the in-plane OOO angle are set to 1.267 Å and 117.2° (cf. Figure \ref{fig:O3_mol}). These spin-orbit integrals are evaluated in cartesian coordinates using the Breit–Pauli Hamiltonian   as implemented in MOLPRO}
    \begin{tabular}{|c|c|c|}
    \hline
      & $\mathrm{R_{OO}} = 1.403$ \AA  & $\mathrm{R_{OO}} = 1.438$ \AA \\
      \hline
      $\bra{X^1\Sigma_g^+}H_{SO}\ket{1^3A''}$ & 17.4i (x) / 7.4 (y) & - \\
      \hline
      $\bra{A^3B_2}H_{SO}\ket{1^3A''}$ & - & 22.8i(x) / 7.6 (y)\\
      \hline
    \end{tabular}
    \label{tab:so}
\end{table}

\end{document}